\documentclass[aps,prl,reprint]{revtex4-2}

\usepackage{lipsum}
\usepackage{amsmath,amssymb}
\usepackage{bm}
\usepackage{physics}
\usepackage{graphicx}
\usepackage[caption=false]{subfig}
\usepackage{diagbox,eqparbox,hhline}
\usepackage{babel}
\usepackage{comment}
\usepackage{soul}
\usepackage{xcolor}
\usepackage{lineno}
\usepackage[colorlinks=true,citecolor=blue,
    linkcolor=blue, urlcolor=blue]{hyperref}
\begin{document}

    \title{
    Transient and universal regimes in quantum reaction-transport kinetics
    }

	\author{Hossein Hosseinabadi}
    \email{hossein@pks.mpg.de}

    \author{Roderich Moessner}
    
    \affiliation{Max Planck Institute for the Physics of Complex Systems, 01187 Dresden, Germany}

	\begin{abstract}
        {Quantum reaction-transport systems consist of coherently propagating particles that irreversibly react upon encounter. Their relaxation is commonly classified as reaction-limited or transport-limited, depending on the relative timescales of reaction and particle transport.} We show that this expectation fails in low dimensions by studying the quantum binary annihilation, $A+A\rightarrow \O$, where reaction processes acquire singular fluctuation corrections below the upper critical dimension $D_c=2$. Consequently, fluctuations dominate the asymptotic kinetics for $D<2$. In one dimension, they render mean-field relaxation transient and drive the system toward a transport-limited regime with $n\sim t^{-1/2}$, governed by a quantum-Zeno scale even for arbitrarily weak loss. At $D=2$, the kinetics acquires logarithmic corrections, whereas above two dimensions mean-field scaling is asymptotically restored. At and below the upper critical dimension, mean-field behavior can nevertheless persist over parametrically long crossover times before the asymptotic fluctuation-dominated regime emerges. We also show that the same fluctuations generate effective elastic collisions despite the absence of microscopic coherent interactions. In $D>1$, these collisions redistribute momentum populations and are parametrically faster than losses in the transport-limited regime, allowing the system to approach a quasi-stationary thermal state.
	\end{abstract}
	
	\maketitle

    Reaction-diffusion phenomena are a paradigmatic subject of nonequilibrium classical statistical physics~\cite{Privman_1997,Hinrichsen,Tauber_2005,delRazo_RMP2026}, describing systems in which particles are transported and react upon encounter, with broad relevance across physics, chemistry, and biology. Their kinetics are commonly classified into reaction-limited and transport-limited regimes, the latter reducing to diffusion-limited kinetics when the underlying motion is diffusive. In the reaction-limited regime, reactions are sufficiently slow for mean-field (MF) kinetics to apply, whereas in the transport-limited regime, the relaxation is controlled by particle motion and fluctuations can modify the dynamics, particularly so in low dimensions~\cite{Mattis_RMP1998,Tauber_2005,Kang_1985}. This distinction gives rise to different universality classes determined by the reaction process and the spatial dimensionality.

    Quantum reaction-transport systems have attracted increasing interest in recent years, motivated in part by advances in cold-atom experiments~\cite{Syassen_2008,Ospelkaus_2010,Ni_2010,Tomita2017,Tomita2019}. Their kinetics has been studied numerically and through analytical approaches based on time-dependent generalized Gibbs ensembles (GGE), kinetic theories, and effective descriptions in different reaction and transport regimes~\cite{vanHorssen_2015,Lange_GGE2018,Rossini_2021,Rosso_2023,Perfetto_2023,Perfetto_PRE2023,Gerbino_2024,Gerbino_2025,Lehr_2025,Gao_PRA2025,Gao_PRL2025}. A general understanding of their universal long-time behavior, however, remains lacking.

    \begin{figure}[!t]
        \centering
        \includegraphics[width=\linewidth]{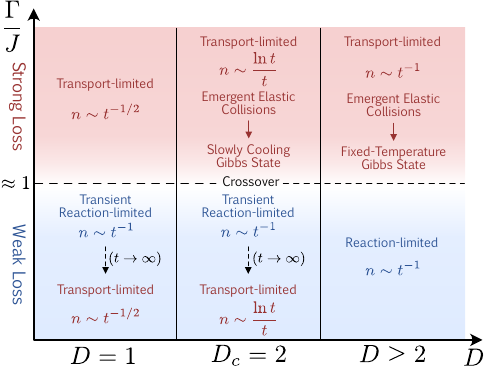}
        \caption{Schematic kinetic-regime diagram as a function of spatial dimension $D$ and loss strength $\Gamma/J$. In $D=1$, the asymptotic dynamics is transport-limited with $n\sim t^{-1/2}$, although weak loss can produce a long transient reaction-limited regime. $D=2$ is marginal, with $n\sim (\ln t)/t$, while for $D > 2$ the mean-field decay $n\sim t^{-1}$ is asymptotically restored. At strong loss and $D>1$, emergent elastic collisions can rapidly redistribute particles and establish a quasi-stationary thermal state.}
        \label{fig:phase_diag}
    \end{figure}

    In this work, one basic question we address is: can -- even weakly -- dissipative quantum dynamics become transport-limited at long times? We show that it can. Using a field-theoretical framework that accounts for the renormalization of reaction processes, we find that the long-time kinetics exhibits an upper critical dimension $D_c=2$, as summarized in Fig.~\ref{fig:phase_diag}. For $D<2$, fluctuations dominate the asymptotic dynamics even for arbitrarily weak microscopic loss. In one dimension, this drives the system toward a transport-limited regime with density scaling distinct from mean field, while the marginal $D=2$ exhibits logarithmic corrections. In both cases, the approach to the fluctuation-dominated regime can be preceded by a parametrically long MF transient. Such reaction-limited transients are naturally captured by time-dependent GGE descriptions~\cite{Lange_GGE2018,Perfetto_2023}, while the infrared renormalization identified here controls the eventual asymptotic dynamics. For $D>2$,
    MF kinetics is recovered. 

    As a further consequence of the same fluctuation corrections, effective elastic collisions emerge between particles despite the absence of microscopic coherent interactions. In $D=1$, kinematic constraints prevent these collisions from redistributing momentum populations, whereas for $D>1$ they can dominate over loss processes in the transport-limited regime and drive the system toward a quasi-stationary Gibbs state.

    \emph{Model.---}
We consider bosons on a $D$-dimensional hypercubic lattice subject to binary annihilation. In the Markovian regime, the density matrix $\rho$ evolves according to the quantum master equation~\cite{Breuer_2007}
\begin{equation}\label{eq:QME}
    \partial_t \rho = -i[H,\rho] + D[\rho],
\end{equation}
    with Hamiltonian $H=-J\sum_{\langle ij\rangle}(\psi^\dagger_{i}\psi_{j}+\mathrm{h.c.})$
    where the bosonic operators satisfy
    $[\psi_{i},\psi^\dagger_{j}]=\delta_{ij}$.
    Binary loss is described by the Lindblad dissipator
    $D[\rho]=\sum_i(L_i\rho L_i^\dagger-\{L_i^\dagger L_i,\rho\}/2)$,
    with jump operators $L_i=\sqrt{\Gamma}\,\psi_{i}^2$ and microscopic reaction rate $\Gamma$. Our universality classification concerns the continuum regime with quadratic dispersion. Lattice band structures can contain additional momentum sectors beyond this limit. Extensions to fermions are discussed below.

   \begin{figure}[!t]
        \centering
        \includegraphics[width=.8\linewidth]{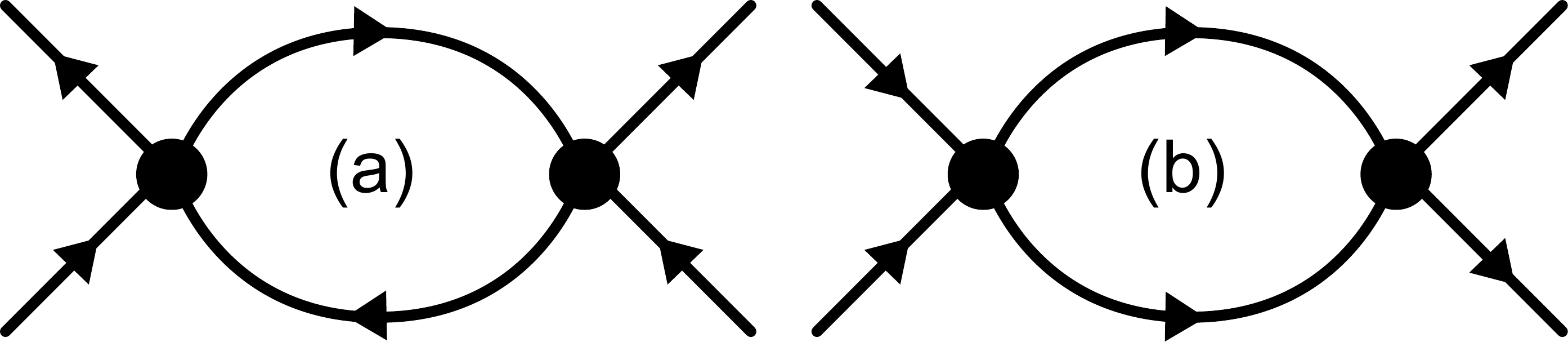}
        \caption{Leading-order corrections to the pair annihilation vertex. Diagram (a) vanishes in the vacuum while (b) survives and is infrared-divergent for $D\le 2$.}
        \label{fig:interaction}
    \end{figure}

    \emph{Method.---}We formulate the problem using Keldysh field theory~\cite{kamenev,Sieberer_2016}, representing the density-matrix evolution on a closed time contour with forward ($+$) and backward ($-$) branches. Introducing bosonic fields on the two branches and performing the Keldysh rotation
    $\psi^{c,q}=(\psi^+\pm\psi^-)/\sqrt{2}$, we obtain the action~\cite{SupplementalMaterial}
    \begin{equation}
        S=S_J+S_\Gamma,
    \end{equation}
    where the free part is
    \begin{equation}
        S_J = \sum_{\bm{k}}\int
        \begin{pmatrix}
            \bar{\psi}^c_{\bm{k}} & \bar{\psi}^q_{\bm{k}}
        \end{pmatrix}
        \begin{pmatrix}
            0 & \qty(G^A_{0\bm{k}})^{-1} \\
            \qty(G^R_{0\bm{k}})^{-1} & 2i0^+
        \end{pmatrix}
        \begin{pmatrix}
            \psi^c_{\bm{k}} \\
            \psi^q_{\bm{k}}
        \end{pmatrix}
        \dd t,
    \end{equation}
    with
    $(G^R_{0\bm{k}})^{-1}=[(G^A_{0\bm{k}})^{-1}]^\dagger
    =i\partial_t-\epsilon_{\bm{k}}+i0^+$ and continuum dispersion
    $\epsilon_{\bm{k}}\approx J\abs{\bm{k}}^2$. The pair-loss contribution is
    \begin{multline}\label{eq:S_Gamma}
    S_\Gamma=\frac{i\Gamma}{2}\sum_i\int \dd t
    \left[
    \qty(\bar{\psi}_{i}^c)^2\psi_{i}^c\psi_{i}^q +\qty(\bar{\psi}_{i}^q)^2\psi_{i}^c\psi_{i}^q \right. \\ \left.
    -\bar{\psi}_i^c\bar{\psi}_i^q \qty(\psi^c_i)^2-\bar{\psi}_i^c\bar{\psi}_i^q \qty(\psi^q_i)^2   - 4 \abs{\psi^c_i}^2\abs{\psi^q_i}^2
    \right],
    \end{multline}
    which acts as a quartic interaction in the Keldysh theory. {Below, we show that the infrared structure responsible for the universal kinetics can already be identified by studying corrections to the reaction of two incoming particles in vacuum.} Readers interested primarily in the long-time kinetics may skip the following diagrammatic derivation and proceed directly to the renormalized kinetic equation, Eq.~\eqref{eq:kinetic}.

       \begin{figure}[!t]
        \centering
        \subfloat[\label{fig:tmatrix}]{\includegraphics[height=0.2\linewidth]{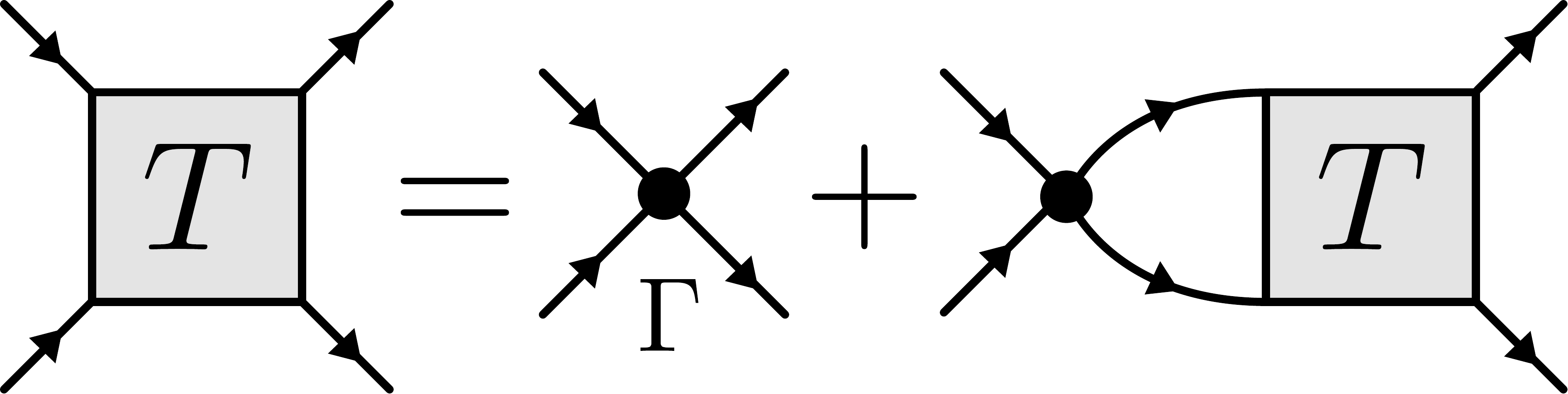}}\vspace{-10pt}\\ \subfloat[\label{fig:selfenergy}]{\includegraphics[height=0.2\linewidth]{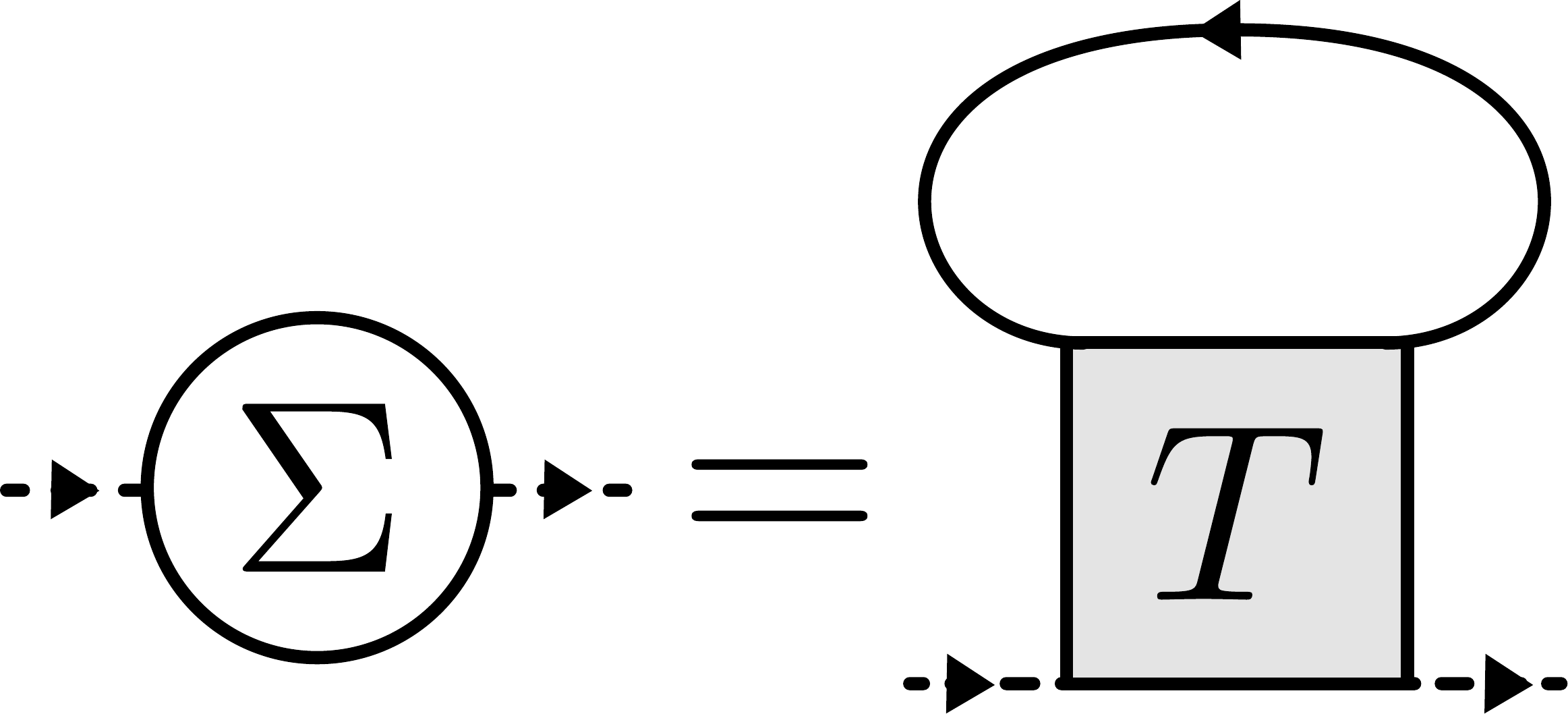}}
        \caption{(a) Self-consistent diagrammatic equation for the T-matrix. (b) Boson self-energy in terms of the T-matrix.}
        \label{fig:diagrams}
        
    \end{figure}
    
    \emph{Perturbation theory.---}We first treat pair loss perturbatively. The second-order corrections to the interaction are shown in Fig.~\ref{fig:interaction}. Diagram (a) involves particle-hole processes and is proportional to the particle density, and therefore vanishes in the vacuum~\cite{SupplementalMaterial}. Diagram (b), by contrast, survives in the vacuum {as it involves virtual particle pair creation. It} depends on the incoming momenta $\bm{q}/2+\bm{p}$ and $\bm{q}/2-\bm{p}$ through
    $\delta\Gamma=-i\Gamma^2\Pi^R(\bm{q}/2+\bm{p},\bm{q}/2-\bm{p})/2$, where $\Pi^R$ is the pair bubble,
    \begin{equation}\label{eq:Pi_def}
        \Pi^R(\frac{\bm{q}}{2}+\bm{p},\frac{\bm{q}}{2}-\bm{p})
        =
        \int
        \frac{2}{E(\bm{q},\bm{p})-E(\bm{q},\bm{k})+i0^+}
        \frac{\dd\bm{k}^D}{(2\pi)^D},
    \end{equation}
    with
    $E(\bm{q},\bm{p})=
    \epsilon_{\bm{q}/2+\bm{p}}+\epsilon_{\bm{q}/2-\bm{p}}$.
    For small $\bm{p}$ and $\bm{q}$, we obtain
    \begin{equation}\label{eq:Pi_scaling}
        \Pi^R(\frac{\bm{q}}{2}+\bm{p},\frac{\bm{q}}{2}-\bm{p})\simeq  \begin{cases}
            -\frac{i}{2J\abs{\bm{p}}},& D=1\\  -\frac{1}{2\pi J}\ln{\frac{\Lambda}{\abs{\bm{p}}}}-\frac{i}{4J}, & D=2\\
            -\frac{\mathcal{A}_D}{J}\qty(\frac{\Lambda^{D-2}}{D-2}+\frac{i\pi}{2}\abs{\bm{p}}^{D-2}), & D\ge 3
        \end{cases}
    \end{equation}
    where $\Lambda$ is a dimensionless ultraviolet cutoff and
    $\mathcal{A}_D^{-1}=2(4\pi)^{D/2}\Gamma(D/2)$.
    The pair bubble is infrared divergent for $D\leq 2$, signaling the breakdown of perturbation theory in low dimensions. The same conclusion follows directly from power counting. At the free fixed point, $z=2$ and the classical and quantum fields have scaling dimensions $[\psi]=D/2$ in the vacuum, giving $[\Gamma]=2-D$. Hence the upper critical dimension is $D_c=2$: reactions are infrared relevant for $D<2$ and marginal at $D=2$.
    
    There are two natural routes beyond perturbation theory. One is a renormalization-group (RG) treatment, in which coarse graining generates additional terms in the action, including coherent quartic interactions of the Hubbard type. The other is the T-matrix approximation~\cite{Nozieres_Tmatrix1985,Semkat_Tmatrix2000}, which resums the most infrared-divergent diagrams at each order in $\Gamma$. {Here we adopt the latter approach and leave the RG formulation for future work.} Besides capturing the nonperturbative renormalization of the reaction vertex, the T-matrix framework provides a direct route to a kinetic description of the dilute finite-density regime, which we use below to analyze the long-time dynamics.

    \emph{Diagrammatic resummation.---}The leading infrared-divergent contributions at higher orders consist of repeated particle-pair bubbles. Resumming the resulting geometric series yields the T-matrix, i.e., the renormalized reaction vertex shown in Fig.~\ref{fig:diagrams}\subref{fig:tmatrix},~\cite{SupplementalMaterial}
    \begin{equation}\label{eq:Tmat_vac}
        T(\bm{k},\bm{p})
        =
        \frac{\Gamma}
        {1+\frac{i}{2}\Gamma\Pi^R(\bm{k},\bm{p})},
    \end{equation}
    which remains finite in all dimensions. Equation~\eqref{eq:Pi_scaling} shows that the infrared divergence of the pair bubble softens the reaction vertex at small $\abs{\bm{k}-\bm{p}}$, with important consequences for the long-time kinetics.
        
    \emph{Renormalized kinetic theory.---}At finite density, reactions break time-translation invariance {as the particle density decays with time,} and the dynamics must be formulated in terms of two-time Green's functions. We introduce the retarded and Keldysh components $G^R_{\bm{k}}(t,t')=-i\Theta(t-t')\langle [\psi_{\bm{k}}(t),\psi_{\bm{k}}^\dagger(t')]\rangle$ and $G^K_{\bm{k}}(t,t')=-i\langle \{\psi_{\bm{k}}(t),\psi_{\bm{k}}^\dagger(t')\}\rangle$, whose equations of motion are~\cite{kamenev}
    \begin{equation}\label{eq:dyson_GR}
        (i\partial_t-\epsilon_{\bm{k}})G^R_{\bm{k}}(t,t')
        =\delta(t-t')+\Sigma^R_{\bm{k}}\otimes G^R_{\bm{k}},
    \end{equation}
    \begin{equation}\label{eq:dyson_GK}
        (i\partial_t-\epsilon_{\bm{k}})G^K_{\bm{k}}(t,t')
        =\Sigma^R_{\bm{k}}\otimes G^K_{\bm{k}}+\Sigma^K_{\bm{k}}\otimes G^A_{\bm{k}},
    \end{equation}
    where $(A\otimes B)(t,t')=\int A(t,t'')B(t'',t')\,\mathrm{d}t''$. Within the T-matrix approximation, the self-energy is shown diagrammatically in Fig.~\ref{fig:diagrams}\subref{fig:selfenergy}, while the T-matrix itself acquires distinct retarded and Keldysh components at finite density. In the dilute regime, where particles propagate approximately freely between rare encounters, these nonlocal equations reduce systematically to a local kinetic equation for the momentum occupations~\cite{SupplementalMaterial}. Importantly, this kinetic theory retains the renormalized reaction vertex and is therefore distinct from a weak-coupling treatment based on the bare rate $\Gamma$.

    The resulting kinetic equation for
    $n_{\bm{k}}=\langle \psi^\dagger_{\bm{k}}\psi_{\bm{k}}\rangle$
    takes the form~\cite{SupplementalMaterial}
    \begin{equation}\label{eq:kinetic}
        \dv{}{t}n_{\bm{k}}
        =
        I_{\mathrm{loss}}[n]
        +
        I_{\mathrm{coll}}[n],
    \end{equation}
    where the loss contribution is
    \begin{equation}\label{eq:I_loss}
        I_{\mathrm{loss}}[n]
        =
        -\frac{4}{\Gamma}
        \int
        \abs{T(\bm{k},\bm{p})}^2
        n_{\bm{k}}n_{\bm{p}}
        \frac{\dd^D\bm{p}}{(2\pi)^D},
    \end{equation}
    and the elastic collision integral is
    \begin{multline}\label{eq:I_coll}
        I_{\mathrm{coll}}[n]
        =
        4\pi
        \iint
        \abs{T(\bm{k},\bm{p})}^2
        \delta(
        \epsilon_{\bm{k}}+\epsilon_{\bm{p}}
        -\epsilon_{\bm{p'}}
        -\epsilon_{\bm{k+p-p'}}
        )
        \\
        \times
        \left(
        n_{\bm{p}'}
        n_{\bm{k+p-p'}}
        -
        n_{\bm{k}}
        n_{\bm{p}}
        \right)
        \frac{\dd^D\bm{p}}{(2\pi)^D}
        \frac{\dd^D\bm{p}'}{(2\pi)^D}.
    \end{multline}
     The appearance of the renormalized vertex $T(\bm{k},\bm{p})$ directly connects the kinetic theory to the vacuum analysis above. Remarkably, effective elastic collisions emerge despite the absence of microscopic coherent interactions: pair loss acts as an imaginary contact potential, suppressing overlap and generating predominantly reflective scattering through the quantum-Zeno effect. Bose-statistics corrections enter only at cubic order in density and are therefore absent from Eq.~\eqref{eq:I_coll}, which is retained  to quadratic order. We now analyze Eq.~\eqref{eq:kinetic} in different spatial dimensions.

    \emph{One-dimensional kinetics.---}The strongest deviation from weak-coupling kinetics occurs in $D=1$, where fluctuations are most pronounced. The collision integral in Eq.~\eqref{eq:I_coll} vanishes because elastic two-body scattering can only exchange particle momenta and therefore leaves the momentum occupations unchanged. Substituting Eq.~\eqref{eq:Pi_scaling} into the T-matrix and subsequently into the loss integral gives
    \begin{equation}\label{eq:kinetic_D1}
        \dv{}{t}n_{k}
        =
        -\int
        \frac{4\Gamma J^2\abs{k-p}^2}
        {\qty(J\abs{k-p}+\Gamma/2)^2}
        \,n_{k}n_{p}\,
        \frac{\dd p}{2\pi}.
    \end{equation}
    The reaction kernel crosses over from the bare rate $4\Gamma$ at large relative momenta to the soft form
    $\propto (J^2/\Gamma)\abs{k-p}^2$
    at small relative momenta. In this regime, stronger loss suppresses reactions while stronger hopping enhances them, signaling transport-limited kinetics. Remarkably, this asymptotic regime emerges in one dimension irrespective of the microscopic loss strength. The resulting scale has the characteristic quantum-Zeno form $J^2/\Gamma$~\cite{Rossini_2021,Rosso_2023,Marche_2024}, familiar from the strong-loss limit where reactions proceed through virtual processes.

    \begin{figure}[!t]
        \centering
        \hspace{-8pt}\subfloat[\label{fig:nk_loc}]{\includegraphics[trim={7pt 0 1pt 0}, clip,height=0.58\linewidth]{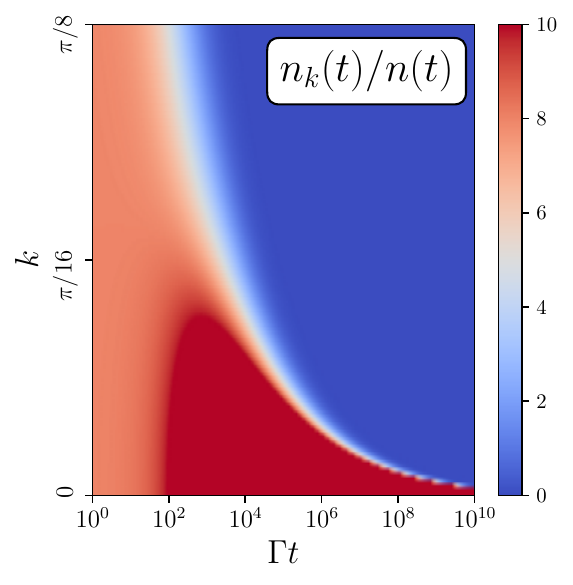}}\hspace{-3pt}\subfloat[\label{fig:n_tot_loc}]{\includegraphics[trim={5pt 0 7pt 0}, clip,height=0.57\linewidth]{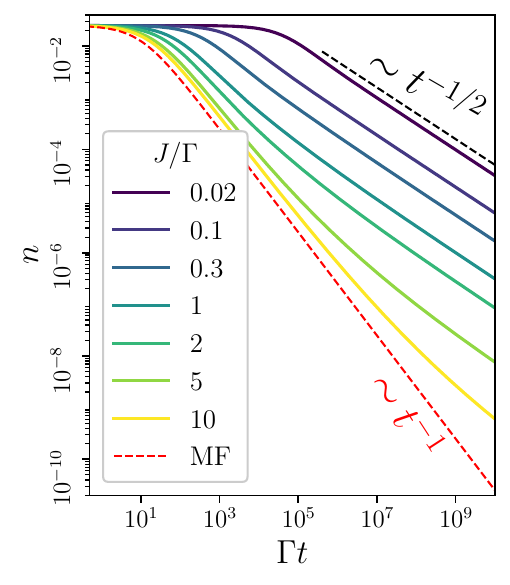}}
        \caption{(a) Time evolution of the momentum distribution normalized by the average density in one dimension for $J=\Gamma$. Higher-momentum modes decay more rapidly, causing the surviving population to concentrate near $k=0$. (b) Density evolution for different hopping strengths. The asymptotic decay follows $n\sim t^{-1/2}$ for all $J/\Gamma$, while for strong hopping a parametrically long transient mean-field regime $n\sim t^{-1}$ can precede the asymptotic scaling.
}
        \label{fig:panel_loc}
    \end{figure}
    
    Figure~\ref{fig:panel_loc}\subref{fig:nk_loc} shows the evolution of the normalized momentum distribution $n_{k}(t)/n(t)$ obtained by numerically solving Eq.~\eqref{eq:kinetic_D1} for $J=\Gamma$, starting from the initial state $n^0_{k}=\Theta(\pi/8-|k|)/5$. Modes at larger momenta decay more rapidly, causing the distribution to progressively concentrate near $k=0$. The corresponding density evolution for several values of $J/\Gamma$ is shown in Fig.~\ref{fig:panel_loc}\subref{fig:n_tot_loc}. At long times, the density obeys
    \begin{equation}\label{eq:n_t_D1}
        n\sim t^{-1/2}, \qquad D=1.
    \end{equation}
    This exponent follows analytically from the quadratic reaction kernel in Eq.~\eqref{eq:kinetic_D1}~\cite{Rosso_2023,Gerbino_2024}. Remarkably, the asymptotic scaling is independent of the microscopic ratio $J/\Gamma$, which instead controls the crossover timescale. For weak loss, mean-field behavior $n\sim t^{-1}$ can persist for parametrically long times before the $t^{-1/2}$ regime emerges. {In particular, the crossover time depends on the initial state. Denoting the characteristic width of the initial momentum distribution by $k_0$, no mean-field transient occurs when $k_0 \lesssim \Gamma/J$, as the system begins directly in the fluctuation-dominated regime. Conversely, for $k_0 \gg \Gamma/J$, the crossover time scales as $t^\star
    \sim e^{ck_0 J/\Gamma}$, where $c$ is a positive dimensionless constant determined by the shape of the initial distribution~\cite{SupplementalMaterial}.}

    \emph{Two-dimensional kinetics.---}At the marginal dimension $D=2$, the pair bubble divergence in Eq.~\eqref{eq:Pi_scaling} is logarithmic. Consequently, for weak loss, fluctuations become important only below the exponentially small momentum scale
    $\abs{\bm{k}-\bm{p}}\sim \exp(-4\pi J/\Gamma)$.
    Elastic collisions are also present, with loss and collision rates scaling as $I_\mathrm{loss}\propto J^2/\Gamma$ and $I_\mathrm{coll}\propto J$. For $J\gg\Gamma$, losses are therefore parametrically faster than collisions, and the latter can be neglected. In this case, starting from a broad class of smooth initial distributions, we obtain at long times~\cite{SupplementalMaterial}
    \begin{equation}\label{eq:n_t_D2}
        n\sim \frac{\ln t}{t},
        \qquad D=2,
    \end{equation}
    i.e., logarithmic corrections to mean-field scaling. In the opposite regime $J\lesssim\Gamma$, elastic collisions dominate over losses and rapidly drive an arbitrary dilute initial distribution toward a quasi-stationary Gibbs state, which is subsequently depleted on a much longer timescale. In this regime we have $n_{\bm{k}}(t)\propto n(t)e^{-J|\bm{k}|^2/T(t)}/T(t)$, with the scaling behavior of density still given by Eq.~\eqref{eq:n_t_D2}, and a time-dependent temperature following~\cite{SupplementalMaterial}
    \begin{equation}
        T(t)\sim J e^{-\sqrt{2\ln(J^2 t/\Gamma)}}.
    \end{equation}
    Therefore, temperature very slowly decreases as the particles are depleted.

    \emph{Kinetics above two dimensions.---}For $D>2$, the T-matrix approaches a constant at small momenta, yielding $\dot n\propto -n^2$. The asymptotic density therefore follows the mean-field scaling
    \begin{equation}\label{eq:n_t_D3}
        n\sim t^{-1},
        \qquad D>2.
    \end{equation}
    In the reaction-limited regime $J\gg\Gamma$, the loss and collision rates scale as
    $I_\mathrm{loss}\propto\Gamma$ and
    $I_\mathrm{coll}\propto\Gamma^2/J$, so elastic collisions are parametrically slower than losses and the dynamics is controlled by the bare reaction rate. In the opposite, transport-limited regime $J\ll\Gamma$,
    $I_\mathrm{loss}\propto J^2/\Gamma$ and
    $I_\mathrm{coll}\propto J$, such that collisions dominate over the Zeno-suppressed losses and drive the system toward a quasi-stationary Gibbs state, while the density continues to obey Eq.~\eqref{eq:n_t_D3}.
    
    In contrast to the two-dimensional case, here the temperature remains constant in the asymptotic low-momentum regime. When elastic collisions are parametrically faster than losses, they rapidly establish a Gibbs state whose temperature is fixed by the initial energy per particle. The subsequent momentum-independent loss preserves this temperature while depleting the density. This difference follows directly from the momentum dependence of the loss kernel. For $D>2$, the kernel approaches a constant, so all momentum modes decay at the same fractional rate,
    $\dot n_{\bm{k}}/n_{\bm{k}}\propto -n$, leaving the normalized momentum distribution unchanged. At $D=2$, by contrast, the loss kernel retains a weak logarithmic momentum dependence, and particles at larger momenta decay more rapidly. The loss thus preferentially removes higher-energy particles, causing the energy to decrease slightly faster than the particle number and leading to a slow cooling.

    \emph{Fermions.---}So far, we have focused on bosons with on-site two-body loss. The same framework extends straightforwardly to spinless fermions, for which on-site two-body loss is forbidden by Fermi statistics and the leading local process involves neighboring sites, $L_i=\sqrt{\Gamma}\,\psi_i\psi_{i+1}$~\cite{Perfetto_2023,Gerbino_2024,Gerbino_2025,Lehr_2025,Gao_PRA2025,Gao_PRL2025}. In one dimension, applying the same T-matrix construction gives the following loss kernel for the kinetic equation~\cite{SupplementalMaterial}
    \begin{equation}\label{eq:Tmat_spinless}
        \frac{1}{\Gamma}|T(k,p)|^2
        \simeq
        \begin{cases}
            \Gamma\abs{k-p}^2, & J\gg\Gamma,\\[4pt]
            \dfrac{16J^2}{\Gamma}\abs{k-p}^2, & J\ll\Gamma.
        \end{cases}
    \end{equation}
    In the reaction-limited regime, this reproduces the weak-loss Hartree-Fock/GGE result in which the reaction rate is set by the bare coupling~\cite{Perfetto_2023,Gerbino_2024}. In the transport-limited regime, by contrast, the rate is governed by the Zeno scale $J^2/\Gamma$. This behavior is consistent with simple power counting: in the continuum limit, the jump operator takes the derivative form $L(x)\sim\psi(x)\partial_x\psi(x)$, giving the scaling dimension $[\Gamma]=-D$. The reaction vertex is therefore infrared irrelevant in all dimensions, so weak-loss perturbation theory remains valid. {In this case, the quadratic dependence on the relative momentum $|k-p|$ in Eq.~\eqref{eq:Tmat_spinless} yields the scaling $n\sim t^{-1/2}$. However, this behavior originates from the interplay of nearest-neighbor pair loss and Fermi statistics, rather than from fluctuation corrections.}

    \emph{Outlook.---}{The present framework suggests several natural extensions.} Above, the upper critical dimension $D_c=2$ followed from the quadratic dispersion underlying the continuum theory. Lifshitz-type band extrema hosting higher-order van Hove singularities, characterized by $\epsilon_{\bm{k}}\propto |\bm{k}|^\alpha$ with $\alpha>2$, can therefore increase the upper critical dimension and generate distinct universal exponents. Such dispersions can be engineered through longer-range hopping, providing a possible route toward realizing new reaction universality classes in cold-atom systems. Lattice effects beyond the continuum limit offer another intriguing direction: additional extrema of the group velocity can become dynamically protected and lead to an initial-state-dependent selection of distinct kinetic regimes. These lattice-induced effects will be addressed elsewhere. Extending the renormalized kinetic theory to spatially inhomogeneous states would also facilitate direct comparison with experiments in trapping potentials. Finally, when binary reactions are forbidden by symmetries or conservation laws, three- and higher-body reactions become the leading processes and may give rise to new nonperturbative regimes of quantum reaction kinetics.

    \begin{acknowledgments}
    \emph{Acknowledgments.---}H.H. is grateful to the Centro de Ciencias de Benasque Pedro Pascual for its hospitality during part of this work. This work was supported in part by the the Deutsche Forschungsgemeinschaft via the cluster of excellence ctd.qmat (EXC 2147, project-id 390858490) and  FOR 5522 (Project-ID No. 499180199).
    \end{acknowledgments}

\bibliography{Refs}

\end{document}


\title{Supplemental Material:\\Transient and universal regimes in quantum reaction-transport kinetics}

\author{Hossein Hosseinabadi}

\author{Roderich Moessner}
\affiliation{Max Planck Institute for the Physics of Complex Systems, 01187 Dresden, Germany}

\maketitle

\section{Non-equilibrium Field Theory}
\subsection{Keldysh action}
Starting from the master equation with
    \begin{equation}\label{eq:QME}
    \partial_t \rho = -i[H,\rho]+\sum_i\left(L_i\rho L_i^\dagger-\frac{1}{2}\left\{L_i^\dagger L_i,\rho\right\}\right),
    \end{equation}
    \begin{equation}\label{eq:H}
    H=-J\sum_{\langle ij\rangle}\,\left(\psi_{i}^\dagger \psi_{j}+\mathrm{h.c.}\right),
    \end{equation}
    \begin{equation}
        L_i=\sqrt{\Gamma}\,\psi_i^2 ,
    \end{equation}
We use the bosonic fields $(\bar{\psi}_{i}^\pm,\psi_{i}^\pm)$ to write down the Keldysh path-integral~\cite{kamenev}. Here $\pm$ specifies the index of the Keldysh contour to which the fields belongs. For the free part of the  action we have
\begin{equation}
    S_J=\sum_{s=\pm}\int \frac{\dd^D \bm{k}}{(2\pi)^D}\int \dd t\, s\bar{\psi}^s_{\bm{k}}(i\partial_t - \epsilon_{\bm{k}})\psi^s_{\bm{k}},
\end{equation}
where $\epsilon_{\bm{k}}=-2J \sum_{\alpha=1}^D \cos{k_\alpha}$ is the single particle dispersion which, at low momenta, becomes $\epsilon_{\bm{k}}\approx J\abs{\bm{k}}^2$. For the dissipative part of the action we follow the prescription of Ref.~\cite{Sieberer_2016} to get
\begin{equation}\label{eq:S_Gamma_pm}
    S_\Gamma = - \frac{i\Gamma}{2} \sum_i \int \mathrm{d}t\, \left(2\qty(\bar{\psi}_{i}^- )^2 \qty(\psi_{i}^+)^2  -  \abs{\psi_i^+}^4 -   \abs{\psi_i^-}^4\right).
\end{equation}
We change the basis and work in terms of the classical and quantum components defined as~\cite{kamenev}
\begin{equation}
    \psi_{i}^c=\frac{1}{\sqrt{2}}\left(\psi_{i}^+ +\psi_{i}^-\right), \quad \psi_{i}^q=\frac{1}{\sqrt{2}}\left(\psi_{i}^+ -\psi_{i}^-\right),
\end{equation}
in terms of which the free action becomes
\begin{equation}\label{eq:S_J}
    S_J= \int \frac{\dd^D \bm{k}}{(2\pi)^D} \int \dd t \,
        \begin{pmatrix}
            \bar{\psi}^c_{\bm{k}} & \bar{\psi}^q_{\bm{k}}
        \end{pmatrix}
        \begin{pmatrix}
            0 & i\partial_t - \epsilon_{\bm{k}}-i0^+ \\
            i\partial_t - \epsilon_{\bm{k}}+i0^+ & 2i0^+
        \end{pmatrix}
        \begin{pmatrix}
            \psi^c_{\bm{k}} \\
            \psi^q_{\bm{k}}
        \end{pmatrix},
\end{equation}
where the infinitesimal regularization for the $qq$ element of the matrix is introduced to enforce the vacuum state. Expansion of $S_\Gamma$ in terms of classical and quantum fields after some algebra leads to
\begin{equation}\label{eq:S_Gamma}
    S_\Gamma=\frac{i\Gamma}{2}\sum_i\int \dd t
    \left[
    \qty(\bar{\psi}_{i}^c)^2\psi_{i}^c\psi_{i}^q +\qty(\bar{\psi}_{i}^q)^2\psi_{i}^c\psi_{i}^q 
    -\bar{\psi}_i^c\bar{\psi}_i^q \qty(\psi^c_i)^2-\bar{\psi}_i^c\bar{\psi}_i^q \qty(\psi^q_i)^2   - 4 \abs{\psi^c_i}^2\abs{\psi^q_i}^2
    \right].
\end{equation}

\subsection{Perturbation theory}
\begin{figure*}
    \centering
    \includegraphics[height=0.08\linewidth]{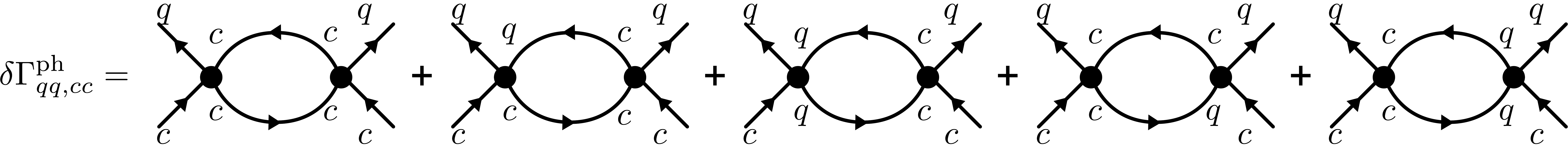}\\ \vspace{10pt}\includegraphics[height=0.08\linewidth]{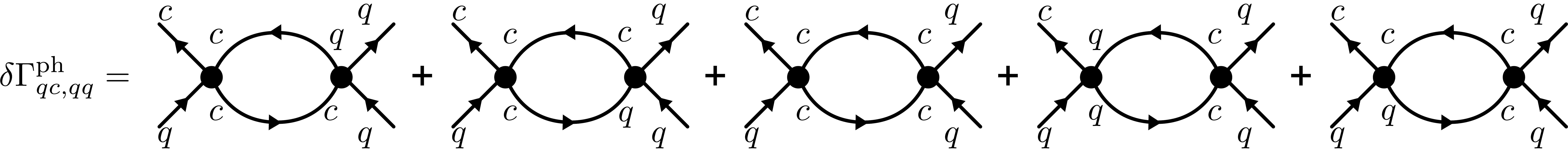}\\ \vspace{10pt}\includegraphics[height=0.08\linewidth]{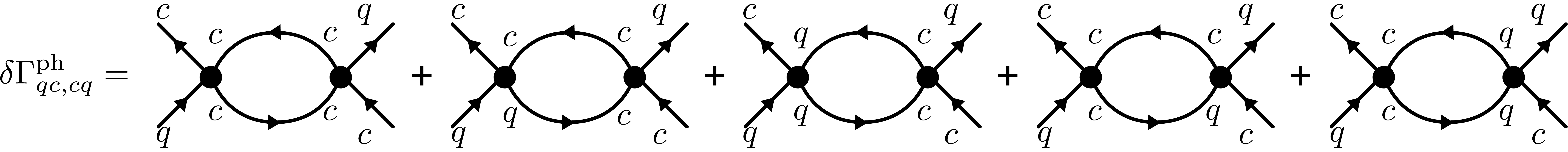}\\ \vspace{10pt}\includegraphics[height=0.08\linewidth]{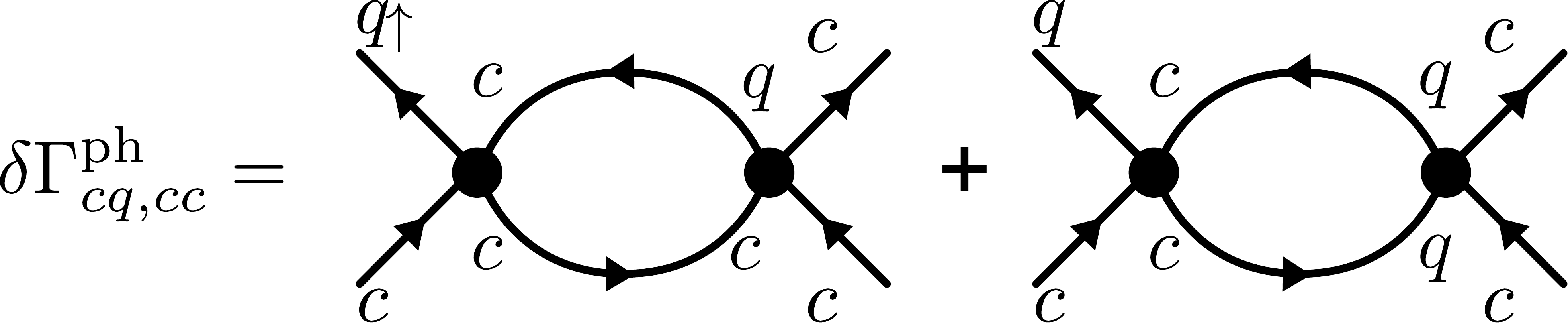}\\ \vspace{10pt}\includegraphics[height=0.08\linewidth]{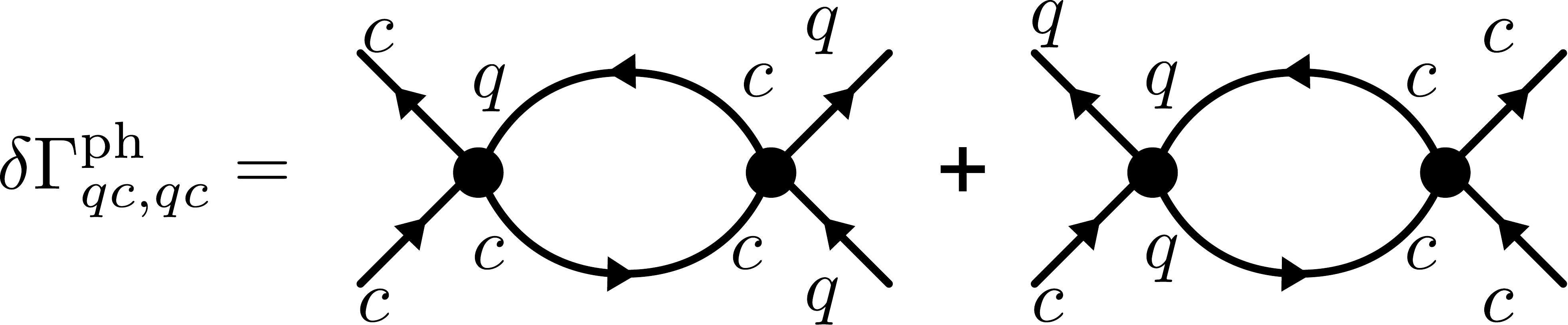}
    \caption{Leading-order correction to the 2-particle vertex in the particle-hole channel. Each row separately vanishes in the vacuum.}
    \label{fig:Gamma2_particlehole}
\end{figure*}

Below, we calculate the corrections to $\Gamma$ to the second-order in perturbation theory. For this purpose, we define the retarded, advanced and Keldysh (symmetric) Green's functions as
\begin{align}
    G^R_{\bm{k}}(t,t')&=-i \expval{\psi^c_{\bm{k}}(t) \bar{\psi}^q_{\bm{k}}(t')},\\
    G^A_{\bm{k}}(t,t')&=-i \expval{\psi^q_{\bm{k}}(t) \bar{\psi}^c_{\bm{k}}(t')},\\
    G^K_{\bm{k}}(t,t')&=-i \expval{\psi^c_{\bm{k}}(t) \bar{\psi}^c_{\bm{k}}(t')}.
\end{align}
In the vacuum, Green's functions are given by their free values since for a single particle, there are no other particles to interact with. We can therefore write $G^{R/K/A}$ in the frequency domain by inverting the kernel in Eq.~\eqref{eq:S_J}
\begin{equation}\label{eq:G_vac}
    G^R(\bm{k},\omega)=\qty(G^A(\bm{k},\omega))^*=\frac{1}{\omega-\epsilon_{\bm{k}}+i0^+}, \qquad G^K(\bm{k},\omega)=G^R(\bm{k},\omega) - G^A(\bm{k},\omega)=-2i\pi \delta(\omega-\epsilon_{\bm{k}}).
\end{equation}
We want to calculate the correction to the 4-point vertices in Eq.~\eqref{eq:S_Gamma}. There are different possible combinations of external legs with different quantum/classical indices. Diagrams in the particle-hole channel (Fig.~2(a) of the main text), have different combinations of classical/quantum indices of their external legs. These are shown in Fig.~\ref{fig:Gamma2_particlehole}, excluding those which are related to these by reversing all arrows, which have the same functional dependence as the displayed diagrams, up to an overall sign. In the vacuum we have $G^K_{\bm{k}}(t-t')=G^R_{\bm{k}}(t-t')-G^A_{\bm{k}}(t-t')$. After substituting this into the mathematical expressions for the diagrams and taking care of all the positive and negative prefactors in Eq.~\eqref{eq:S_Gamma}, one can see that all terms in Fig.~\ref{fig:Gamma2_particlehole} vanish. For instance:
\begin{equation}
    \delta \Gamma_{qc,qc}^\mathrm{ph}(\bm{k},t)\propto -\Gamma^2 \int\frac{\dd^D \bm{p}}{(2\pi)^D} \qty[ G^A_{\bm{p+k}}(t)G^K_{\bm{p}}(-t) + G^K_{\bm{p+k}}(t)G^R_{\bm{p}}(-t) + 2G^A_{\bm{p+k}}(t)G^R_{\bm{p}}(-t)]=0.
\end{equation}
Physically, the cancellation occurs due to the fact that all intermediate processes in Fig.~\ref{fig:Gamma2_particlehole} involve the application of the annihilation operator to the vacuum.

\begin{figure*}
    \centering
    \includegraphics[height=0.08\linewidth]{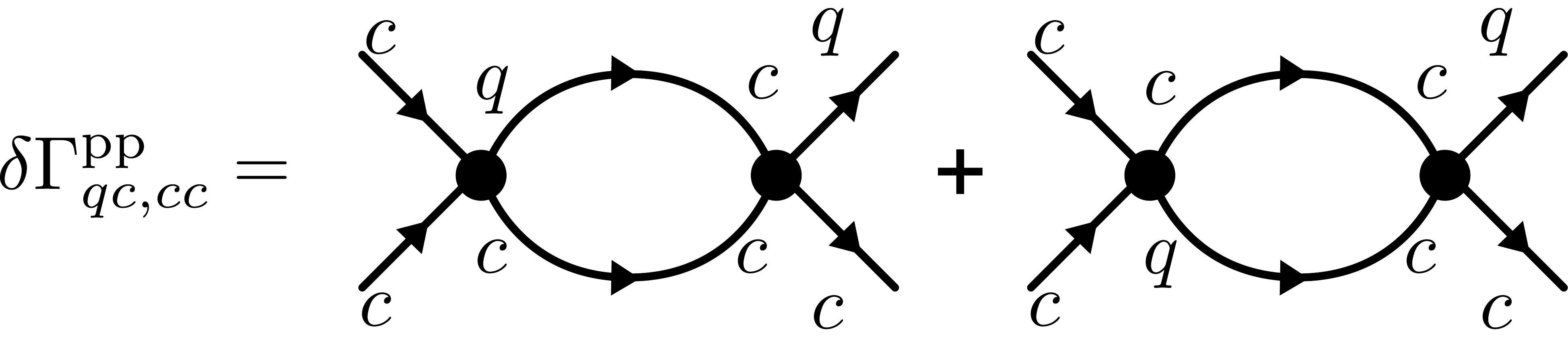}\hspace{40pt}\includegraphics[height=0.08\linewidth]{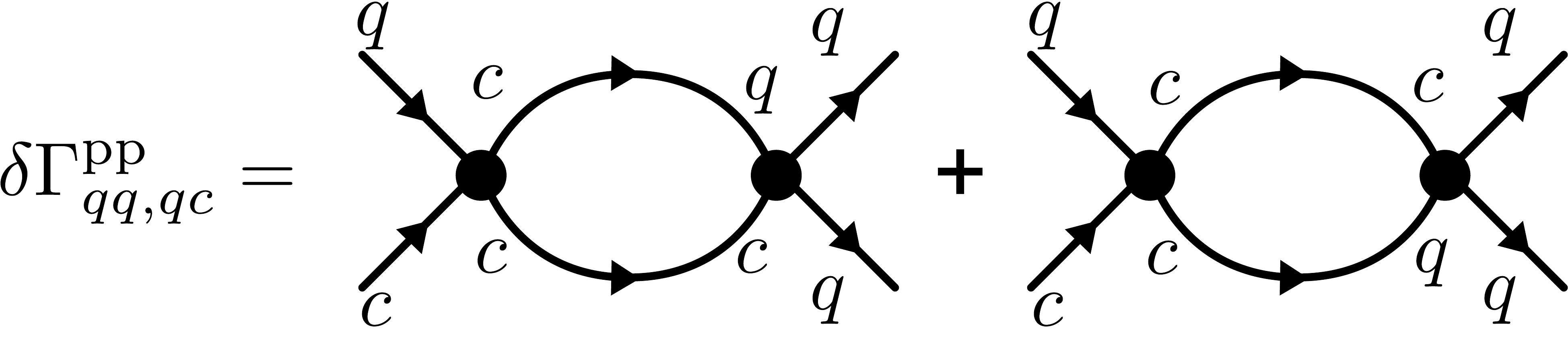}\\ \vspace{10pt}\includegraphics[height=0.08\linewidth]{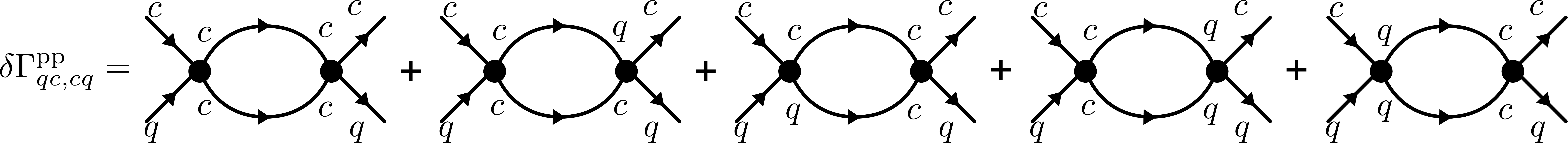}
    \caption{Leading-order correction to the 2-particle vertex in the pairing channel.}
    \label{fig:Gamma2_pp}
\end{figure*}

However, the corrections in the particle-particle (pairing) channel do not vanish. These are shown in Fig.~\ref{fig:Gamma2_pp}. Using the vacuum condition $G^K_{\bm{k}}(t-t')=G^R_{\bm{k}}(t-t')-G^A_{\bm{k}}(t-t')$ and the general property of retarded and advanced functions $G^R(t)G^A(t)=0$, one can show that all of them are proportional to the retarded pair bubble $\Pi^R_{\bm{q}}(t)$ (or its complex conjugate) defined as
\begin{equation}
    \Pi^R_{\bm{q}}(t)=2i \int G^R_{\bm{q}/2+\bm{p}}(t)G^K_{\bm{q}/2-\bm{p}}(t)\, \frac{\dd^D \bm{p}}{(2\pi)^D}.
\end{equation}
In the frequency space we get
\begin{equation}\label{eq:PiR_freq}
    \Pi^R(\bm{q},\omega)=2i\int\int G^R(\bm{q}/2+\bm{p},\omega/2+\nu)G^K(\bm{q}/2-\bm{p},\omega/2-\nu)\, \frac{\dd^D \bm{p}}{(2\pi)^D}\frac{\dd \nu}{2\pi}.
\end{equation}
Using Eq.~\eqref{eq:G_vac} we get
\begin{equation}
    \Pi^R(\bm{q},\omega)=\int \frac{2}{\omega-\epsilon_{\bm{q}/2+\bm{k}}-\epsilon_{\bm{q}/2-\bm{k}}+i0^+}\, \frac{\dd^D \bm{p}}{(2\pi)^D}.
\end{equation}
In the final step, we also note that the frequency $\omega$ is the total energy of two incoming particles with center of mass momentum $\bm{q}$. Taking the momenta of the two incoming particles as $\bm{q}/2\pm\bm{p}$, we have $\omega=\epsilon_{\bm{q}/2+\bm{p}}+\epsilon_{\bm{q}/2-\bm{p}}$. Upon substitution into the equation above we obtain Eq.~(5) of the main text.

\subsection{Self-consistent T-matrix approximation (SCTA)}
The essence of SCTA is to resum all of the (divergent) pair bubble diagrams to infinite order~\cite{Nozieres_Tmatrix1985,Semkat_Tmatrix2000}. While this can be done iteratively by directly calculating higher-order diagrams, in a similar manner to the previous section, a more convenient approach is to apply a Hubbard-Stratonovich transformation to Eq.~\eqref{eq:S_Gamma} in the pairing channel. This introduces a new complex auxiliary field $\phi^{c,q}_{\bm{q}}$ which is coupled to bosonic bilinears. The result is a mapping $S_\Gamma \to S_\phi + S_{\psi \phi}$ where
\begin{equation}
    S_\chi = \int \frac{\dd^D \bm{k}}{(2\pi)^D}\int \dd t\,
        \bar{\Phi}^T_{\bm{k}}(t)\cdot \hat{T}^{-1}_0\cdot \Phi_{\bm{k}}(t), \quad \Phi_{\bm{k}}(t)=\begin{pmatrix}
            \phi^c_{\bm{k}}(t) & \phi^q_{\bm{k}}(t)
        \end{pmatrix}, \quad \hat{T}^{-1}_0=\frac{2i}{\Gamma}\begin{pmatrix}
            0 & -1 \\ +1 & +2
        \end{pmatrix},
\end{equation}
and
\begin{equation}
    S_{\psi \phi} = \frac{1}{\sqrt{2}}\iint \frac{\dd^D \bm{k}}{(2\pi)^D} \frac{\dd^D \bm{p}}{(2\pi)^D} \int \dd t\,\qty[2\bar{\phi}_{\bm{k}+\bm{p}}^c\psi^c_{\bm{k}}\psi^q_{\bm{p}}+\bar{\phi}_{\bm{k}+\bm{p}}^q\qty(\psi^c_{\bm{k}}\psi^c_{\bm{p}}+\psi^q_{\bm{k}}\psi^q_{\bm{p}}) + \mathrm{c.c.}].
\end{equation}
We can check that evaluating the Gaussian integral over $\phi$ yields $S_\Gamma$. We then define the T-matrix as the correlation functions of the auxiliary field
\begin{align}
    T^R_{\bm{k}}(t,t')&=-i \expval{\phi^c_{\bm{k}}(t) \bar{\phi}^q_{\bm{k}}(t')},\\T^A_{\bm{k}}(t,t')&=-i \expval{\phi^q_{\bm{k}}(t) \bar{\phi}^c_{\bm{k}}(t')},\\T^K_{\bm{k}}(t,t')&=-i \expval{\phi^c_{\bm{k}}(t) \bar{\phi}^c_{\bm{k}}(t')}.
\end{align}
Our goal is to obtain the equations of motion governing $G^{R/A/K}$ and $T^{R/A/K}$. For bosons, these are the Schwinger-Dyson equations quoted in Eqs.~(8) and (9) of the main text
\begin{align}
        (i\partial_t-\epsilon_{\bm{k}})G^R_{\bm{k}}(t,t')
        &=\delta(t-t')+\Sigma^R_{\bm{k}}\otimes G^R_{\bm{k}},\label{eq:dyson_GR}\\(i\partial_t-\epsilon_{\bm{k}})G^A_{\bm{k}}(t,t')
        &=\delta(t-t')+\Sigma^A_{\bm{k}}\otimes G^A_{\bm{k}},\label{eq:dyson_GA}\\
        (i\partial_t-\epsilon_{\bm{k}})G^K_{\bm{k}}(t,t')
        &=\Sigma^R_{\bm{k}}\otimes G^K_{\bm{k}}+\Sigma^K_{\bm{k}}\otimes G^A_{\bm{k}},\label{eq:dyson_GK},
    \end{align}
and for the T-matrix we have
\begin{equation}\label{eq:dyson_TR}
    T^R_{\bm{k}}(t,t')=-\frac{i\Gamma}{2}\delta(t-t')-\frac{i\Gamma}{2}\int \Pi^R_{\bm{k}}(t,t'')T^R_{\bm{k}}(t'',t')\, \dd t'',
\end{equation}
\begin{equation}
    T^A_{\bm{k}}(t,t')=+\frac{i\Gamma}{2}\delta(t-t')+\frac{i\Gamma}{2}\int \Pi^A_{\bm{k}}(t,t'')T^A_{\bm{k}}(t'',t')\, \dd t'',
\end{equation}
\begin{equation}\label{eq:dyson_TK}
    T^K_{\bm{k}}(t,t')=-2T^A_{\bm{k}}(t,t')-\frac{i\Gamma}{2}\int \qty[ \Pi^R_{\bm{k}}(t,t'')T^K_{\bm{k}}(t'',t')+\Pi^K_{\bm{k}}(t,t'')T^A_{\bm{k}}(t'',t')]\, \dd t''.
\end{equation}
The functions $\Sigma^{R/A/K}$ and $\Pi^{R/A/K}$ are the self-energies for bosons and the auxiliary field. Their full evaluation is equivalent to solving the problem exactly, which in general is not possible. Within SCTA, we approximate these at the one-loop level, whose Feynman diagrams were shown in Fig.~3 of the main text,
\begin{equation}
    \Sigma^R_{\bm{k}}(t,t')= \qty(\Sigma^A_{\bm{k}}(t',t))^*=2i \int \qty[ T^R_{\bm{k}+\bm{p}}(t,t') G_{\bm{p}}^K(t',t)  + T^K_{\bm{k}+\bm{p}}(t,t') G_{\bm{p}}^A(t',t) ]\,\frac{\dd^D \bm{p}}{(2\pi)^D},
\end{equation}
\begin{equation}
    \Sigma^K_{\bm{k}}(t,t')=2i \int \qty[ T^K_{\bm{k}+\bm{p}}(t,t') G_{\bm{p}}^{\bm{K}}(t',t) + T^R_{\bm{k}+\bm{p}}(t,t') G_{\bm{p}}^A(t',t)   + T^A_{\bm{k}+\bm{p}}(t,t') G_{\bm{p}}^R(t',t) ] \,\frac{\dd^D \bm{p}}{(2\pi)^D}.
\end{equation}
The self-energy for the auxiliary field is given by
\begin{equation}
    \Pi^R_{\bm{q}}(t,t')=2i \int G^R_{\bm{q}/2+\bm{k}}(t,t')G^K_{\bm{q}/2-\bm{k}}(t,t') \, \frac{\dd^D \bm{k}}{(2\pi)^D},
\end{equation}
\begin{equation}
    \Pi^K_{\bm{q}}(t,t')=i \int\left[ G^K_{\bm{k}}(t,t')G^K_{\bm{q}-\bm{k}}(t,t') + G^R_{\bm{k}}(t,t')G^R_{\bm{q}-\bm{k}}(t,t') + G^A_{\bm{k}}(t,t')G^A_{\bm{q}-\bm{k}}(t,t')\right]  \, \frac{\dd^D \bm{p}}{(2\pi)^D}.
\end{equation}
It can be shown that $\Sigma$ and $\Pi$ can be obtained from a Luttinger-Ward functional in terms of the Green's functions. An important consequence of this is that the resulting approximation is conserving, meaning that it respects the conservation laws of the problem. Since the system is dissipative, the total particle number and energy are not conserved. However, in cases where the dissipation becomes effectively weak, such as the collision-dominated regimes discussed in the text, the conserving approximation ensures that the system relaxes to a quasi-stationary thermal ensemble. 

In the vacuum state, we can go to the frequency domain where Eq.~\eqref{eq:dyson_TR} gives
\begin{equation}
    T^R(\bm{q},\omega)=\qty(T^A(\bm{q},\omega))^*=\frac{-i\Gamma/2}{1+i\Gamma\Pi^R(\bm{q},\omega)/2},
\end{equation}
where $\Pi^R(\bm{q},\omega)$ is given by Eq.~\eqref{eq:PiR_freq}. Eq.~\eqref{eq:dyson_TR} suggests that the renormalized value of $\Gamma$ should be identified with $2i T^R(\bm{q},\omega)$ upon putting $\omega$ equal to the total energy of an incoming pair, which leads to Eq.~(7) of the main text.

\section{Kinetic theory in the dilute limit}

The equations of motion derived above depend on two time variables, making their numerical solution at long times computationally expensive. In the dilute regime, however, they reduce to a local kinetic theory~\cite{kamenev}. We introduce the Wigner transform
\begin{equation}
    f(t,\omega)
    =
    \int \dd\tau\,
    e^{i\omega\tau}
    f\left(t+\frac{\tau}{2},t-\frac{\tau}{2}\right).
\end{equation}
In the dilute limit, the retarded Green's function
$G^R_{\bm{k}}(t,\omega)$ is given, to leading order in the density, by its vacuum value. We therefore expand the Keldysh component around the vacuum solution,
\begin{equation}\label{eq:GK}
    G^K_{\bm{k}}(t,\omega)
    \approx
    G^K_{0}(\bm{k},\omega)
    +
    \delta G^K_{\bm{k}}(t,\omega),
\end{equation}
where
\begin{equation}\label{eq:GK0}
    G^K_{0}(\bm{k},\omega)
    =
    G^R_{0}(\bm{k},\omega)
    -
    G^A_{0}(\bm{k},\omega)
    =
    -2\pi i\,\delta(\omega-\epsilon_{\bm{k}}).
\end{equation}
The occupation $n_{\bm{k}}(t)$ is encoded in
$\delta G^K_{\bm{k}}$ according to
\begin{equation}\label{eq:deltaGK}
    \delta G^K_{\bm{k}}(t,\omega)
    =
    2
    \left[
    G^R_{0}(\bm{k},\omega)
    -
    G^A_{0}(\bm{k},\omega)
    \right]
    n_{\bm{k}}(t)
    =
    -4\pi i\,n_{\bm{k}}(t)
    \delta(\omega-\epsilon_{\bm{k}}).
\end{equation}
Applying the Wigner transformation to
Eq.~\eqref{eq:dyson_GK} and its Hermitian conjugate,
and subtracting the resulting equations, gives
\begin{equation}\label{eq:dt_deltaGK}
    \partial_t\delta G^K_{\bm{k}}(t,\omega)
    =
    2\Im\Sigma^R_{\bm{k}}(t,\omega)
    \left[
    G^K_{0}(\bm{k},\omega)
    +
    \delta G^K_{\bm{k}}(t,\omega)
    \right]
    +
    i\Sigma^K_{\bm{k}}(t,\omega)
    G^K_{0}(\bm{k},\omega).
\end{equation}
The bosonic self-energies after the Wigner transformation are
\begin{equation}
\Sigma^R_{\bm{k}}(t,\omega)
=
2i
\int\frac{\dd^D \bm{p}}{(2\pi)^D}
\int\frac{\dd\nu}{2\pi}
\Big[
T^R_{\bm{k}+\bm{p}}(t,\omega+\nu)
\left(
G^K_{0}(\bm{p},\nu)
+
\delta G^K_{\bm{p}}(t,\nu)
\right)
+
T^K_{\bm{k}+\bm{p}}(t,\omega+\nu)
G^A_{0}(\bm{p},\nu)
\Big],
\label{eq:SigR}
\end{equation}
and
\begin{equation}
\Sigma^K_{\bm{k}}(t,\omega)
=
2i
\int\frac{\dd^D \bm{p}}{(2\pi)^D}
\int\frac{\dd\nu}{2\pi}
\Big[
T^K_{\bm{k}+\bm{p}}(t,\omega+\nu)
\left(
G^K_{0}(\bm{p},\nu)
+
\delta G^K_{\bm{p}}(t,\nu)
\right)
-
\left(
T^R_{\bm{k}+\bm{p}}(t,\omega+\nu)
-
T^A_{\bm{k}+\bm{p}}(t,\omega+\nu)
\right)
G^K_{0}(\bm{p},\nu)
\Big].
\label{eq:SigK}
\end{equation}
For the retarded and advanced T-matrices,
\begin{equation}\label{eq:TR}
    T^R_{\bm{q}}(t,\omega)
    =
    \left(T^A_{\bm{q}}(t,\omega)\right)^*
    =
    \frac{-i\Gamma/2}
    {1+i\Gamma\Pi^R_{\bm{q}}(t,\omega)/2},
\end{equation}
while the Keldysh component is
\begin{equation}\label{eq:TK}
    T^K_{\bm{q}}(t,\omega)
    =
    \frac{
    \Gamma^2\Pi^K_{\bm{q}}(t,\omega)/4-i\Gamma
    }
    {\left|
    1+i\Gamma\Pi^R_{\bm{q}}(t,\omega)/2
    \right|^2}.
\end{equation}
In the vacuum,
$T^K_0=T^R_0-T^A_0$, which motivates the decomposition
\begin{equation}\label{eq:TK_decomp}
    T^K_{\bm{q}}(t,\omega)
    \equiv
    T^R_{\bm{q}}(t,\omega)
    -
    T^A_{\bm{q}}(t,\omega)
    +
    \Delta T^K_{\bm{q}}(t,\omega).
\end{equation}
Using Eqs.~\eqref{eq:TR} and~\eqref{eq:TK}, we obtain
\begin{equation}
    \Delta T^K_{\bm{q}}(t,\omega)
    =
    \frac{\Gamma^2}{4}
    \frac{
    \Delta\Pi^K_{\bm{q}}(t,\omega)
    }
    {\left|
    1+i\Gamma\Pi^R_{\bm{q}}(t,\omega)/2
    \right|^2},
\end{equation}
where
\begin{equation}
    \Delta\Pi^K_{\bm{q}}(t,\omega)
    \equiv
    \Pi^K_{\bm{q}}(t,\omega)
    -
    \left[
    \Pi^R_{\bm{q}}(t,\omega)
    -
    \Pi^A_{\bm{q}}(t,\omega)
    \right].
\end{equation}
The retarded pair propagator is
\begin{equation}
    \Pi^R_{\bm{q}}(t,\omega)
    =
    2i
    \int\frac{\dd^D \bm{k}}{(2\pi)^D}
    \int\frac{\dd\nu}{2\pi}\,
    G^R_{0}(\bm{k},\nu)
    G^K_{\bm{q}-\bm{k}}
    (t,\omega-\nu).
\end{equation}
Substituting Eqs.~\eqref{eq:GK}-\eqref{eq:deltaGK},
we write
\begin{equation}
    \Pi^R_{\bm{q}}(t,\omega)
    =
    \Pi_0^R(\bm{q},\omega)
    +
    \delta\Pi^R_{\bm{q}}(t,\omega),
\end{equation}
with the vacuum contribution
\begin{equation}\label{eq:PiR0}
    \Pi^R_0(\bm{q},\omega)
    =
    2\int\frac{\dd^D \bm{k}}{(2\pi)^D}
    \frac{1}
    {\omega-\epsilon_{\bm{k}}
    -\epsilon_{\bm{q}-\bm{k}}+i0^+},
\end{equation}
and
\begin{equation}
    \delta\Pi^R_{\bm{q}}(t,\omega)
    =
    4\int\frac{\dd^D \bm{k}}{(2\pi)^D}
    \frac{
    n_{\bm{q}-\bm{k}}(t)
    }
    {\omega-\epsilon_{\bm{k}}
    -\epsilon_{\bm{q}-\bm{k}}+i0^+}.
\end{equation}
A similar expansion of $\Pi^K$ gives
\begin{equation}\label{eq:DeltaPiK}
    \Delta\Pi^K_{\bm{q}}(t,\omega)
    =i\int\frac{\dd^D \bm{k}}{(2\pi)^D}
    \int\frac{\dd\nu}{2\pi}\,
    \delta G^K_{\bm{k}}(t,\nu)
    \delta G^K_{\bm{q}-\bm{k}}
    (t,\omega-\nu),
\end{equation}
which is quadratic in the density. We now substitute Eq.~\eqref{eq:TK_decomp} into
Eqs.~\eqref{eq:SigR} and~\eqref{eq:SigK} and retain
terms up to quadratic order in the occupations. This gives
\begin{equation}
\Sigma^R_{\bm{k}}(t,\omega)
\simeq
2i
\int\frac{\dd^D \bm{p}}{(2\pi)^D}
\int\frac{\dd\nu}{2\pi}
\Big[
T^R_{\bm{k}+\bm{p}}(t,\omega+\nu)
\delta G^K_{\bm{p}}(t,\nu)
+
\Delta T^K_{\bm{k}+\bm{p}}(t,\omega+\nu)
G^A_{0}(\bm{p},\nu)
\Big],
\end{equation}
and
\begin{equation}
\Sigma^K_{\bm{k}}(t,\omega)
\simeq
2i
\int\frac{\dd^D \bm{p}}{(2\pi)^D}
\int\frac{\dd\nu}{2\pi}
\Big[
\left(
T^R_{\bm{k}+\bm{p}}(t,\omega+\nu)
-
T^A_{\bm{k}+\bm{p}}(t,\omega+\nu)
\right)
\delta G^K_{\bm{p}}(t,\nu)
+
\Delta T^K_{\bm{k}+\bm{p}}(t,\omega+\nu)
G^K_{0}(\bm{p},\nu)
\Big].
\end{equation}
It is useful to write
\begin{equation}
    \Sigma^K_{\bm{k}}
    =
    \Sigma^R_{\bm{k}}
    -
    \Sigma^A_{\bm{k}}
    +
    \Delta\Sigma^K_{\bm{k}}.
\end{equation}
With the conventions above,
\begin{equation}
    \Delta\Sigma^K_{\bm{k}}(t,\omega)
    =
    4i
    \int\frac{\dd^D \bm{p}}{(2\pi)^D}
    \int\frac{\dd\nu}{2\pi}\,
    G^K_{0}(\bm{p},\nu)
    \Delta T^K_{\bm{k}+\bm{p}}(t,\omega+\nu),
\end{equation}
which is second order in the density.
Substituting this relation into
Eq.~\eqref{eq:dt_deltaGK}, the terms linear in the density
cancel, yielding
\begin{equation}
    i\partial_t
    \delta G^K_{\bm{k}}(t,\omega)
    =
    2i\Im\Sigma^R_{\bm{k}}(t,\omega)
    \delta G^K_{\bm{k}}(t,\omega)
    -
    G^K_{0}(\bm{k},\omega)
    \Delta\Sigma^K_{\bm{k}}(t,\omega).
\end{equation}
Using Eqs.~\eqref{eq:GK0} and~\eqref{eq:deltaGK}
and integrating over $\omega$ puts the external particles
on shell, $\omega=\epsilon_{\bm{k}}$, giving
\begin{equation}\label{eq:dt_nk}
    \partial_t n_{\bm{k}}(t)
    =
    2n_{\bm{k}}(t)
    \Im\Sigma^R_{\bm{k}}
    (t,\epsilon_{\bm{k}})
    +
    \frac{i}{2}
    \Delta\Sigma^K_{\bm{k}}
    (t,\epsilon_{\bm{k}}).
\end{equation}
To quadratic order in the density, only the term linear in
$\delta G^K$ is needed in $\Im\Sigma^R$, giving
\begin{equation}\label{eq:Im_SigR}
\Im\Sigma^R_{\bm{k}}
(t,\epsilon_{\bm{k}})
=
4\int\frac{\dd^D \bm{p}}{(2\pi)^D}\,
\Im T^R_{0}(\bm{k}+\bm{p},\epsilon_{\bm{k}}+\epsilon_{\bm{p}})
n_{\bm{p}}(t)
=
-2\Gamma
\int\frac{\dd^D \bm{p}}{(2\pi)^D}
\frac{
1-\frac{\Gamma}{2}
\Im\Pi^R_{0}(\bm{k},\bm{p}
)
}
{
\left|
1+\frac{i\Gamma}{2}
\Pi^R_{0}(\bm{k},\bm{p})
\right|^2
}
n_{\bm{p}}(t),
\end{equation}
where we have changed the notation $\Pi^R_{0}(\bm{k}+\bm{p}
,\epsilon_{\bm{k}}+\epsilon_{\bm{p}}) \to \Pi^R_{0}(\bm{k},\bm{p})$ after putting $\omega=\epsilon_{\bm{k}}+\epsilon_{\bm{p}}$. Similarly,
\begin{equation}\label{eq:Delta_SigK}
    \Delta\Sigma^K_{\bm{k}}
    (t,\epsilon_{\bm{k}})
    =
    i\Gamma^2
    \int\frac{\dd^D \bm{p}}{(2\pi)^D}
    \frac{
    \Im\Delta\Pi^K_{\bm{k}+\bm{p}}
    (t,\epsilon_{\bm{k}}+\epsilon_{\bm{p}})
    }
    {
    \left|
    1+\frac{i\Gamma}{2}
    \Pi^R_{0}(\bm{k},\bm{p})
    \right|^2
    }.
\end{equation}

The imaginary part of the vacuum pair propagator is
\begin{equation}\label{eq:Im_PiR0}
    \Im\Pi_0^R(\bm{q},\omega)
    =
    -2\pi
    \int\frac{\dd^D \bm{k'}}{(2\pi)^D}\,
    \delta\left(
    \omega-\epsilon_{\bm{k'}}
    -\epsilon_{\bm{q}-\bm{k'}}
    \right),
\end{equation}
while Eq.~\eqref{eq:DeltaPiK} gives
\begin{equation}\label{eq:Im_DeltaPiK}
    \Im\Delta\Pi^K_{\bm{q}}(t,\omega)
    =
    -8\pi
    \int\frac{\dd^D \bm{k'}}{(2\pi)^D}\,
    n_{\bm{k'}}(t)
    n_{\bm{q}-\bm{k'}}(t)
    \delta\left(
    \omega-\epsilon_{\bm{k'}}
    -\epsilon_{\bm{q}-\bm{k'}}
    \right).
\end{equation}
Substituting these into Eqs.~\eqref{eq:Im_SigR}~and~\eqref{eq:Delta_SigK} and then into Eq.~\eqref{eq:dt_nk}, we get Eq.~(10)-(12) of the main text.

\section{Estimation of the Mean-field scaling transient in one dimension}
The emergence of the asymptotic scaling at long times in $D=1$ is a direct consequence of the fact that the distribution width decreases with time, as shown in Fig.~4(a) of the main text. Eventually, when the width becomes smaller than $\Gamma/J$, the soft nature of the reaction kernel becomes relevant. The shrinkage of the width itself originates from the faster decay of particles at higher momenta as they encounter more particles. We therefore expand the loss kernel as
\begin{equation}
    \frac{4\Gamma J^2 |k-p|^2}{(J|k-p|+\Gamma/2)^2}\simeq 4\Gamma \qty(1-\frac{\Gamma}{J|k-p|}),
\end{equation}
where the first term is the mean-field value and the second term is the leading-order correction at large $J$. For the fastest decaying modes near the edges of the momentum distribution, characterized by $Q(t)$, we can approximately write
\begin{equation}\label{eq:n_edge_n_avg}
    \frac{\dot{n}_\mathrm{edge}}{n_\mathrm{edge}} - \frac{\dot{n}_\mathrm{avg}}{n_\mathrm{avg}} \sim - \frac{\Gamma^2}{J Q}n_\mathrm{avg}.
\end{equation}
This quantity itself should be proportional to the relative shrinking rate of the distribution
\begin{equation}
    \frac{\dot{Q}}{Q} \sim \frac{\dot{n}_\mathrm{edge}}{n_\mathrm{edge}} - \frac{\dot{n}_\mathrm{avg}}{n_\mathrm{avg}}.
\end{equation}
Using Eq.~\eqref{eq:n_edge_n_avg} we obtain
\begin{equation}
    \dot{Q}\sim - \frac{\Gamma^2}{J}n.
\end{equation}
Using the mean-field scaling $n_\mathrm{avg}\sim 1/\Gamma t$, and integrating over time we get
\begin{equation}
    Q-k_0\sim -\frac{\Gamma}{J}\ln(\frac{t}{t_0}),
\end{equation}
where $k_0$ is proportional to the initial width. Mean-field scaling persists until $Q\simeq \Gamma/J$, assuming that $k_0 \gg \Gamma/J$, we obtain
\begin{equation}
    t \sim \exp(\frac{c k_0 J}{\Gamma}).
\end{equation}

\section{Asymptotic kinetics in two dimensions}
\label{sec:D2_kinetics}

We now derive the long-time density scaling in two dimensions. 
The important feature of $D=2$ is the logarithmic infrared divergence of the vacuum pair propagator. 
On shell, and at small relative momentum, it takes the form
\begin{equation}
    \Pi^R_0(\bm{k},\bm{p})
    \simeq
    -\frac{1}{2\pi J}
    \ln\left(\frac{2\Lambda}{|\bm{k}-\bm{p}|}\right)
    -\frac{i}{4J},
\end{equation}
Deep in the infrared,
\begin{equation}
    \frac{\Gamma}{4\pi J}
    \ln\left(\frac{2\Lambda}{|\bm{k}-\bm{p}|}\right)
    \gg 1,
\end{equation}
the corresponding loss kernel becomes
\begin{equation}
    \mathcal K(|\bm{k}-\bm{p}|)
    \equiv
    4\frac{| T(\bm{k},\bm{p})|^2}{\Gamma}
    \simeq
    \frac{\kappa}
    {\ln^2\left(2\Lambda/|\bm{k}-\bm{p}|\right)},
    \qquad
    \kappa\equiv\frac{64\pi^2J^2}{\Gamma}.
    \label{eq:D2_kernel}
\end{equation}
The logarithmic softening of Eq.~\eqref{eq:D2_kernel} is responsible for the marginal correction to mean-field kinetics. We now show that the same asymptotic density scaling,
\begin{equation}
    n(t)\sim\frac{\ln t}{t},
    \label{eq:D2_scaling_SM}
\end{equation}
is obtained both when losses dominate over elastic collisions and when collisions rapidly establish a thermal state.

\subsection{Loss-dominated regime: $\Gamma\ll J$}

For $\Gamma\ll J$, the characteristic collision and loss rates satisfy
$I_{\rm coll}/I_{\rm loss}\sim\Gamma/J\ll1$, so the collision integral can be neglected asymptotically. Taking the kinetic equation
\begin{equation}
    \partial_t n_{\bm{k}}
    =
    -n_{\bm{k}}
    \int\frac{\dd^2\bm{p}}{(2\pi)^2}
    \mathcal K(|\bm{k}-\bm{p}|)
    n_{\bm{p}},
    \label{eq:D2_loss_only}
\end{equation}
we consider a broad class of smooth initial distributions whose long-time evolution approaches a single-scale form,
\begin{equation}
    n_{\bm{k}}(t)
    =
    \frac{n(t)}{Q^2(t)}
    F\left(\frac{\bm{k}}{Q(t)}\right),
    \label{eq:D2_scaling_ansatz}
\end{equation}
where
\begin{equation}
    \int\frac{\dd^2\bm{x}}{(2\pi)^2}F(\bm{x})=1,
    \qquad
    \int\frac{\dd^2\bm{x}}{(2\pi)^2}
    |\bm{x}|^2F(\bm{x})=1.
\end{equation}
With this convention,
\begin{equation}
    Q^2(t)
    =
    \frac{1}{n(t)}
    \int\frac{\dd\bm{k}^2}{(2\pi)^2}
    |\bm{k}|^2 n_{\bm{k}}(t)
\end{equation}
measures the width of the momentum distribution. Introducing
\begin{equation}
    L(t)\equiv\ln\left(\frac{2\Lambda}{Q(t)}\right),
\end{equation}
and writing $\bm{k}=Q\bm{x}$ and $\bm{p}=Q\bm{y}$, the loss kernel becomes
\begin{equation}
    \mathcal K(Q|\bm{x}-\bm{y}|)
    =
    \frac{\kappa}
    {\left(
    L-\ln|\bm{x}-\bm{y}|
    \right)^2}.
\end{equation}
For $L\gg1$, its average over any regular scaling function with finite logarithmic moments can be expanded as
\begin{equation}
    \mathcal K(Q|\bm{x}-\bm{y}|)
    =
    \frac{\kappa}{L^2}
    \left[
    1+
    \frac{2\ln|\bm{x}-\bm{y}|}{L}
    +O(L^{-2})
    \right].
    \label{eq:D2_kernel_expansion}
\end{equation}
Although this expansion is not pointwise uniform at $\bm{x}=\bm{y}$, it is well defined after integration since in two dimensions
$\int_0^1 r\,\dd r\,|\ln r|^m<\infty$ for any finite $m$. Integrating Eq.~\eqref{eq:D2_loss_only} over momentum gives
\begin{equation}
    \dot n
    =
    -\frac{\kappa n^2}{L^2}
    \left[
    1+\frac{2C_0}{L}
    +O(L^{-2})
    \right],
    \label{eq:D2_n_flow}
\end{equation}
where
\begin{equation}
    C_0
    =
    \int
    F(\bm{x})F(\bm{y})
    \ln|\bm{x}-\bm{y}|\,\frac{\dd^2\bm{x}}{(2\pi)^2}\frac{\dd^2\bm{y}}{(2\pi)^2}.
\end{equation}
To determine the evolution of $Q$, we consider the second moment
\begin{equation}
    M_2
    \equiv
    \int \frac{\dd^2\bm{k}}{(2\pi)^2}|\bm{k}|^2n_{\bm{k}}
    =
    nQ^2.
\end{equation}
Using Eq.~\eqref{eq:D2_loss_only},
\begin{equation}\label{eq:dot_M2}
    \dot M_2
    =
    -\frac{\kappa n^2Q^2}{L^2}
    \left[
    1+\frac{2C_2}{L}
    +O(L^{-2})
    \right],
\end{equation}
with
\begin{equation}
    C_2
    =
    \int
    F(\bm{x})F(\bm{y})
    |\bm{x}|^2\ln|\bm{x}-\bm{y}|\,\frac{\dd^2\bm{x}}{(2\pi)^2}\frac{\dd^2\bm{y}}{(2\pi)^2}.
\end{equation}
Using $M_2=nQ^2$, together with Eqs.~\eqref{eq:D2_n_flow}~and~\eqref{eq:dot_M2} yields
\begin{equation}
    \frac{\dot Q}{Q}
    =
    -c\,\frac{\kappa n}{L^3}
    +O\left(\frac{\kappa n}{L^4}\right),
    \qquad
    c\equiv C_2-C_0.
    \label{eq:D2_Q_flow}
\end{equation}
The leading $L^{-2}$ contribution cancels because it is independent of momentum and therefore removes all momentum modes at the same fractional rate. The first momentum-dependent correction appears at order $L^{-3}$ and preferentially depletes particles farther from the center of the distribution. For a regular isotropic profile one therefore has $c>0$. As an explicit example, for a Gaussian profile normalized such that $\langle |\bm{x}|^2\rangle=1$, one obtains $c=1/4$. Since $L=\ln(2\Lambda/Q)$, Eq.~\eqref{eq:D2_Q_flow} gives
\begin{equation}
    \dot L
    =
    c\,\frac{\kappa n}{L^3}.
\end{equation}
Combining this with the leading part of Eq.~\eqref{eq:D2_n_flow},
\begin{equation}
    \frac{\dot n}{n}
    =
    -\frac{\kappa n}{L^2},
\end{equation}
we obtain
\begin{equation}
    \frac{\dd L}{\dd\ln n}
    =
    -\frac{c}{L},
\end{equation}
and hence
\begin{equation}
    L^2
    =
    2c\ln\left(\frac{n_*}{n}\right)
    +O\left(\sqrt{\ln\frac{n_*}{n}}\right),
    \label{eq:D2_L_n}
\end{equation}
where $n_*$ is a nonuniversal constant.
The density equation therefore reduces asymptotically to
\begin{equation}
    \dot n
    \simeq
    -\frac{\kappa}{2c}
    \frac{n^2}
    {\ln(n_*/n)}.
    \label{eq:D2_n_log_eq}
\end{equation}
Defining $x=1/n$, Eq.~\eqref{eq:D2_n_log_eq} gives, to leading logarithmic accuracy,
\begin{equation}
    \dot x
    \simeq
    \frac{\kappa}
    {2c\ln x}.
\end{equation}
Integration yields
\begin{equation}
    t
    \simeq
    \frac{2c}{\kappa}
    x\ln x,
\end{equation}
or equivalently
\begin{equation}
    n(t)
    \simeq
    \frac{2c}{\kappa t}
    \ln(\frac{\kappa t}{2c}) \to n(t)\sim\frac{\ln t}{t}.
\end{equation}
The detailed scaling function $F$ affects the nonuniversal coefficient $c$, but not the logarithmic correction to the density decay.

\subsection{Collision-dominated regime: $\Gamma\gg J$}

For $\Gamma\gg J$, elastic collisions are parametrically faster than losses. They therefore rapidly drive the system toward a Gibbs state and subsequently maintain thermal equilibrium while the much slower loss process depletes the particles. In the dilute regime, the momentum distribution is Maxwellian,
\begin{equation}
    n_{\bm{k}}(t)
    =4\pi J
    \frac{ n(t)}{T(t)}
    \exp\left[-\frac{Jk^2}{T(t)}\right].
    \label{eq:D2_thermal_dist}
\end{equation}
This distribution has precisely the single-scale form introduced in Eq.~\eqref{eq:D2_scaling_ansatz}. Defining
\begin{equation}
    Q^2(t)=\frac{T(t)}{J},
\end{equation}
we can write
\begin{equation}
    n_{\bm{k}}(t)
    =
    \frac{n(t)}{Q^2(t)}
    F\left(\frac{\bm{k}}{Q(t)}\right),
    \qquad
    F(\bm{x})=4\pi e^{-x^2},
    \label{eq:D2_Maxwell_F}
\end{equation}
which satisfies
\begin{equation}
    \int F(\bm{x})\,\frac{\dd^2\bm{x}}{(2\pi)^2}=1,
    \qquad
    \int \abs{\bm{x}}^2F(\bm{x})\,\frac{\dd^2\bm{x}}{(2\pi)^2}=1.
\end{equation}
The analysis of the previous subsection can therefore be applied directly. In particular, the evolution of the characteristic momentum scale is governed by Eq.~\eqref{eq:D2_Q_flow},
\begin{equation}
    \frac{\dot Q}{Q}
    =
    -c\,\frac{\kappa n}{L^3}
    +O\left(\frac{\kappa n}{L^4}\right),
    \qquad
    L=\ln\left(\frac{2\Lambda}{Q}\right),
\end{equation}
where
\begin{equation}
    c=C_2-C_0.
\end{equation}
For the Maxwell distribution in Eq.~\eqref{eq:D2_Maxwell_F}, the integrals can be evaluated explicitly and give
\begin{equation}
    c=\frac14.
\end{equation}
Equation~\eqref{eq:D2_L_n} then becomes
\begin{equation}
    L^2
    \simeq
    \frac12
    \ln\left(\frac{n_*}{n}\right).
    \label{eq:D2_L_n_thermal}
\end{equation}
Consequently, the density equation takes the asymptotic form
\begin{equation}
    \dot n
    \simeq
    -2\kappa
    \frac{n^2}{\ln(n_*/n)},
\end{equation}
and hence
\begin{equation}
    n(t)\sim\frac{\ln t}{t}.
\end{equation}
Thus the rapid elastic collisions modify the shape of the momentum distribution, fixing it to a Maxwell distribution, but do not change the marginal density scaling obtained in the loss-dominated regime. The same result also determines the temperature evolution. Since $Q^2(t)=T/J$ and $L=\ln(2\Lambda/Q)$, we have
\begin{equation}
    T\propto J e^{-2L}.
\end{equation}
Using Eq.~\eqref{eq:D2_L_n_thermal}, this gives
\begin{equation}
    T(n)
    \sim
    J
    \exp\left[
        -\sqrt{
            2\ln\left(\frac{n_*}{n}\right)
        }
    \right].
    \label{eq:D2_T_n}
\end{equation}
Together with $n(t)\sim(\ln J^2t/\Gamma)/(J^2t/\Gamma)$, we obtain
\begin{equation}
    T(t)
    \sim
    J
    e^{-\sqrt{
            2\ln(\frac{J^2 t}{\Gamma})
                +\cdots}}.
    \label{eq:D2_T_time}
\end{equation}
as quoted in the main text.

\section{Asymptotic kinetics above two dimensions}
\label{sec:Dgt2_kinetics}
For $D>2$, the pair propagator is infrared-finite. Using the notation of the main text,
\begin{equation}
    \Pi^R(\bm{k},\bm{p})
    \longrightarrow
    \Pi^R_*,
    \qquad
    |\bm{k}|,|\bm{p}|\rightarrow0,
\end{equation}
where $\Pi^R_*\propto J^{-1}$ is a finite, nonuniversal constant. Consequently, the renormalized reaction vertex approaches a momentum-independent value,
\begin{equation}
    T(\bm{k},\bm{p})
    \longrightarrow
    T_*
    \equiv
    \frac{\Gamma}{
        1+\frac{i\Gamma}{2}\Pi^R_*
    },
\end{equation}
and the asymptotic loss kernel becomes
\begin{equation}
    \mathcal K(\bm{k},\bm{p})
    \equiv
    4\frac{|T(\bm{k},\bm{p})|^2}{\Gamma}
    \longrightarrow
    \mathcal K_*
    \equiv
    4\frac{|T_*|^2}{\Gamma}.
    \label{eq:Kstar_Dgt2}
\end{equation}
The momentum dependence of the reaction kernel is therefore irrelevant at the infrared fixed point. Then, the loss contribution to the kinetic equation reduces to
\begin{equation}
    \partial_t n_{\bm{k}}
    \big|_{\rm loss}
    =
    -\mathcal K_*\,n\,n_{\bm{k}},
    \label{eq:Dgt2_mode_loss}
\end{equation}
where
\begin{equation}
    n=\int\frac{\dd^D\bm{k}}{(2\pi)^D}n_{\bm{k}}.
\end{equation}
Since the collision integral conserves particle number, integration of the full kinetic equation over momentum gives
\begin{equation}
    \dot n
    =
    -\mathcal K_* n^2.
    \label{eq:Dgt2_density_eq}
\end{equation}
Thus
\begin{equation}
    n(t)
    =
    \frac{n_0}{
        1+\mathcal K_*n_0t
    }
    \sim t^{-1},
    \qquad D>2.
    \label{eq:Dgt2_density}
\end{equation}
The mean-field density exponent is therefore asymptotically restored above the upper critical dimension.

\subsection{Quasi-equilibrium in the collision-dominated regime}

In the collision-dominated regime, the fast collision integral pushes the system to the Maxwell distribution
\begin{equation}
    n_{\bm{k}}(t)
    =
    n(t)
    \left(
        \frac{4\pi J}{T(t)}
    \right)^{D/2}
    \exp\left[
        -\frac{Jk^2}{T(t)}
    \right].
    \label{eq:Dgt2_Maxwell}
\end{equation}
The collision integral vanishes on this distribution by detailed balance and subsequently acts only to maintain thermal equilibrium. The crucial difference from $D=2$ is that the asymptotic loss kernel in Eq.~\eqref{eq:Kstar_Dgt2} is independent of momentum. Equation~\eqref{eq:Dgt2_mode_loss} therefore implies
\begin{equation}
    \frac{\partial_t n_{\bm{k}}}{n_{\bm{k}}}
    =
    -\mathcal K_* n,
    \label{eq:Dgt2_fractional_loss}
\end{equation}
which is the same for every momentum mode. Hence the normalized momentum distribution does not evolve under the loss dynamics,
\begin{equation}
    \partial_t
    \left(
        \frac{n_{\bm{k}}}{n}
    \right)
    =0.
    \label{eq:Dgt2_shape_arrest}
\end{equation}
Once the Maxwell distribution has been established, its width therefore remains fixed. The same result can be expressed in terms of the energy density,
\begin{equation}
    \mathcal E
    =
    \int\frac{\dd^D\bm{k}}{(2\pi)^D}
    \epsilon_{\bm{k}}n_{\bm{k}}
    =
    \frac{D}{2}nT.
    \label{eq:Dgt2_energy}
\end{equation}
Because the loss rate is momentum independent,
\begin{equation}
    \dot{\mathcal E}
    =
    -\mathcal K_* n\mathcal E,
\end{equation}
whereas Eq.~\eqref{eq:Dgt2_density_eq} gives
\begin{equation}
    \dot n
    =
    -\mathcal K_*n^2.
\end{equation}
It follows immediately that
\begin{equation}
    \frac{\dd}{\dd t}
    \left(
        \frac{\mathcal E}{n}
    \right)
    =0,
\end{equation}
and therefore
\begin{equation}
    \boxed{
        \dot T=0
    }.
    \label{eq:Dgt2_T_arrest}
\end{equation}
Thus the temperature becomes arrested once the system has reached the asymptotic thermal regime. Since elastic collisions conserve both energy and particle number, the temperature established during the rapid thermalization stage is fixed by the energy per particle before appreciable loss occurs. Denoting this temperature by $T_\infty$,
\begin{equation}
    T_\infty
    =
    \frac{2}{D}
    \frac{\mathcal E_{\rm th}}{n_{\rm th}}
    \simeq
    \frac{2}{D}
    \frac{\mathcal E_0}{n_0},
    \label{eq:Dgt2_Tinf}
\end{equation}
where the last relation applies when the separation between the collision and loss timescales is parametrically large. The subsequent evolution is therefore
\begin{equation}
    n_{\bm{k}}(t)
    =
    n(t)
    \left(
        \frac{4\pi J}{T_\infty}
    \right)^{D/2}
    e^{-Jk^2/T_\infty}.
\end{equation}

\section{Spinless fermions}
\label{sec:spinless}

We finally consider the case of spinless fermions in one dimension. Since two identical fermions cannot undergo an on-site reaction, the shortest-range binary loss process acts on neighboring sites,
\begin{equation}
    L_i=\sqrt{\Gamma}\,\psi_i\psi_{i+1}.
\end{equation}
In momentum space, antisymmetry of the incoming two-particle state generates a momentum-dependent reaction vertex. In one dimension, the corresponding pair propagator is
\begin{equation}
    \Pi^R(q/2+p,q/2-p)
    =
    \int\frac{\dd k}{2\pi}
    \frac{
        2\sin^2(k)
    }{
        \epsilon_{q/2+p}+\epsilon_{q/2-p}-\epsilon_{q/2+k}-\epsilon_{q/2-k}+i0^+
    }.
    \label{eq:Pi_spinless}
\end{equation}
The integral can be evaluated to get
\begin{equation}\label{eq:PiR0_fin}
    \Pi^R(k,p)
    =
    -\frac{
        \left|\cos\left(\frac{k-p}{2}\right)\right|
        +i\left|\sin\left(\frac{k-p}{2}\right)\right|
    }{
        \left|2J\cos\left(\frac{k+p}{2}\right)\right|
    }.
\end{equation}
The corresponding renormalized loss kernel is
\begin{equation}
    \frac{1}{\Gamma}|T(k,p)|^2
    =
    \frac{
        4\Gamma
        \sin^2\left(\frac{k-p}{2}\right)
    }{
        \left|
        1+\frac{i\Gamma}{2}\Pi^R(k,p)
        \right|^2
    }.
    \label{eq:spinless_kernel}
\end{equation}
Using Eq.~\eqref{eq:PiR0_fin}, this can be written explicitly as
\begin{equation}
    \frac{1}{\Gamma}|T(k,p)|^2
    =
    \frac{
        4\Gamma
        \sin^2\left(\frac{k-p}{2}\right)
    }{
        1+
        \dfrac{\Gamma
        \left|\sin\left(\frac{k-p}{2}\right)\right|}
        {2J\left|\cos\left(\frac{k+p}{2}\right)\right|}
        +
        \dfrac{\Gamma^2}
        {16J^2\cos^2\left(\frac{k+p}{2}\right)}
    }.
    \label{eq:spinless_kernel_exact}
\end{equation}
In the continuum limit $|k|,|p|\ll1$, this reduces to
\begin{equation}
    \frac{1}{\Gamma}|T(k,p)|^2
    \simeq
    \frac{
        \Gamma|k-p|^2
    }{
        1+
        \dfrac{\Gamma|k-p|}{4J}
        +
        \dfrac{\Gamma^2}{16J^2}
    }.
    \label{eq:spinless_kernel_cont}
\end{equation}
Consequently,
\begin{equation}
    \frac{1}{\Gamma}|T(k,p)|^2
    \simeq
    \begin{cases}
        \Gamma|k-p|^2,
        & \Gamma\ll J,\\[5pt]
        \displaystyle
        \frac{16J^2}{\Gamma}|k-p|^2,
        & \Gamma\gg J,
    \end{cases}
    \label{eq:spinless_kernel_limits}
\end{equation}
quoted in Eq.~(18) of the main text.

\bibliography{Refs}